\documentstyle[multicol,prb,aps,epsf]{revtex}
\begin{document}
\title{Discrete Hubbard-Stratonovich transformations for systems with
orbital degeneracy}

\author  {O. Gunnarsson}
\address{Max-Planck-Institut f\"ur Festk\"orperforschung, D-70506 Stuttgart,
Germany}
\author{E. Koch}
\address{Department of Physics, University of Illinois,   
          Urbana, Illinois 61801}  

 \date{\today}
\maketitle
\begin{abstract}
A discrete Hubbard-Stratonovich transformation is presented
for systems with an orbital degeneracy $N$ and a Hubbard 
Coulomb interaction without multiplet effects. An exact 
transformation is obtained by introducing an external 
field which takes $N+1$ values. Alternative approximate 
transformations are presented, where the field takes
fewer values, for instance two values corresponding to an
Ising spin.
\end{abstract}
\begin{multicols}{2}
Systems with orbital degeneracy have recently attracted much interest.
For instance, in systems with colossal magnetoresistance, like the
manganites, the degenerate $3d$ orbital plays a major role.
Another example is the alkali-doped C$_{60}$ compounds, 
A$_3$C$_{60}$ (A= K, Rb), where the partly filled $t_{1u}$
orbital has a three-fold degeneracy. Other well-known 
examples are the $3d$ metals. 

Quantum Monte Carlo, auxiliary field methods are popular  
methods for treating strongly correlated systems.           
In this approach a Hubbard-Stratonovich transformation is used
to convert the many-body problem into a one-body problem, at 
the cost of introducing a fluctuating auxiliary 
field.\cite{Hubbard} A substantial simplification was 
introduced by Hirsch, who showed that for a system without 
orbital degeneracy the field only needs to take two values
and that it can be described by an Ising spin.\cite{Hirsch}
This method can be generalized to a system
with the orbital degeneracy $N$ by introducing an Ising
spin for each pair of orbitals. This leads, however,  to a large 
number ($2N(2N-1)/2$) of Ising spins for each site. 
This approach is very general, and can, for
instance, handle the case when multiplet effects  are considered.
In many cases, however, we are interested in models
which only have a Hubbard Coulomb interaction $U$, without
any multiplet effects. This simplifies the problem 
substantially, since the Coulomb energy then only depends
on the total number of electrons on a given site,
and not on the precise occupancy of the different levels.
It is then natural to focus on the occupation
number operator $n$ for a given site. To calculate the
partition function, we would then like to find an 
expression of the type
\begin{equation}\label{eq:1}
e^{- U\Delta \tau n(n-1)/2}=
\sum_{k=1}^{N+1} w_ke^{-x_k n}\equiv \sum_{k=1}^{N+1} w_k z_k^n, 
\end{equation}
where $n=0, 1,..., 2N$ 
corresponding to the possible occupancies of a given site. 
$\Delta \tau$ is the step length in $\tau$ introduced in
the Trotter procedure and $z_k={\rm exp}(-x_k)$.
Because there are $2N+1$ conditions, it should be possible
to satisfy these conditions exactly by letting $k$ run over 
$N+1$ values, since there are then $2N+2$ parameters.
Below we find that this is indeed the case. For $U<0$,
i.e. an attractive interaction, $w_k$ is real and positive 
and $x_k$ is real. For $U>0$ the  $x_k$'s are
complex, but come in pairs of complex conjugates. 

This problem can be solved analytically by rewriting the 
identity used by Hubbard\cite{Hubbard} 
\begin{equation}\label{eq:2}
e^{un(n-1)/2}\equiv \int _{-\infty}^{\infty}e^{-\pi x^2+
(\sqrt{2\pi u}x-u/2)n}dx,
\end{equation}
as
\begin{equation}\label{eq:3}
e^{un(n-1)/2}\equiv \int w(z)z^ndz,   
\end{equation}      
where the weight function $w(z)$ is defined appropriately and 
$u=-U\Delta \tau$.
Since the integral (\ref{eq:3}) now contains a polynomial of a maximum
power $2N$, the integral can be replaced by a discrete sum of the type
in Eq.~(\ref{eq:1}) by using the theory of Gau\ss ian integration
and orthogonal polynomials.\cite{Gauss} The transformation (\ref{eq:1}) is then
exact for $0\le n\le 2N+1$.
Below we reformulate this procedure in such a way that it 
can be generalized to a more useful form.

We introduce a polynomial
\begin{equation}\label{eq:4}
P(z)=\Pi_{k=1}^{N+1}(z-z_k)\equiv\sum_{n=0}^{N+1} \alpha_{n} z^{N+1-n},
\end{equation}
where $\alpha_0=1$. We also define
\begin{equation}\label{eq:5}
B(n)=e^{-U\Delta \tau n(n-1)/2}.
\end{equation}
which is the left hand side of Eq.~(\ref{eq:1}).
From Eq.~(\ref{eq:1}) together with $P(z_k)=0$  we then obtain
\begin{equation}\label{eq:6}
\sum_{m=1}^{N+1} A_{n,m}\alpha_m=C_n      \hskip1cm n=1,..,K,
\end{equation}
where $K=N+1$ for the moment, $A_{n,m}=B(N+n-m)$ and $C_n=-B(N+n)$. 
The solution of Eq.~(\ref{eq:6}) gives  
$\alpha_m$. These coefficients define $P(z)$ from which we obtain the
roots $z_k$. The first $N+1$ equations in Eq.~(\ref{eq:1}) 
then form a set of linear equations in $w_k$, which is solved.
This approach
exactly reproduces the result of the Gau\ss ian method mentioned above.
Finally we transform from $z_k$ to $x_k$ by using $x_k=-{\rm ln} (z_k)$.
Since $B(n)$ is real, the coefficients $\alpha_m$ are also real.
The roots $z_k$ of the polynomial $P(z)$ are then real or come
in pairs of complex conjugates. The same is true for $x_k$.
 For $U<0$ $w_k$ is positive and $x_k$ is real, while 
for $U>0$ $w_k$ and $x_k$ in general come in pairs of complex conjugates. 
In the following we consider a positive (repulsive) $U$.

By using $K=N+1$ in Eq.~(\ref{eq:6}) we really impose a physically 
irrelevant condition, since it means that Eq.~(\ref{eq:1}) is satisfied
for the occupancy $n=2N+1$, which never occurs in the  problem.
We therefore relax this condition and use $K=N$, thereby satisfying
Eq.~(\ref{eq:1}) exactly for $n=0, ..., 2N$. Thus we put $A_{N+1,m}=
\delta_{N+1,m}$. The value of $C_{N+1}\equiv \alpha_{N+1}$ can then
be used to impose some additional condition on the transformation 
in Eq.~(\ref{eq:1}), as is discussed below.

To  use these results in a Monte Carlo approach, we introduce   
\begin{eqnarray}\label{eq:7}
&& p_k={|w_k| \over \sum_j |w_j|}    \\
&& y_k={w_k \over |w_k|}\sum_j |w_j|, \nonumber   \\
\end{eqnarray}
where $p_k\ge 0$ and $\sum_k p_k=1$.
If $w_k$ are real and positive $y_k=1$, since  $\sum_j w_j=1$.
This leads to      
\begin{equation}\label{eq:8}
B(n)\equiv e^{-U\Delta \tau n(n-1)/2}=\sum_{k=1}^{N+1}p_k y_ke^{-x_kn}.       
\end{equation}
We now use a probability interpretation and choose the 
term $y_k e^{-x_k n}$ with the probability $p_k$. 
We introduce the relative standard deviation
\begin{equation}\label{eq:9}
\sigma(n)^2=\sum_{k=1}^{N+1}p_k|{y_ke^{-x_k n }\over B(n)}-1|^2.
\end{equation}

\begin{figure}[bt]
\unitlength1cm
\begin{minipage}[t]{8.5cm}
\centerline{\epsfxsize=3.375in \epsffile{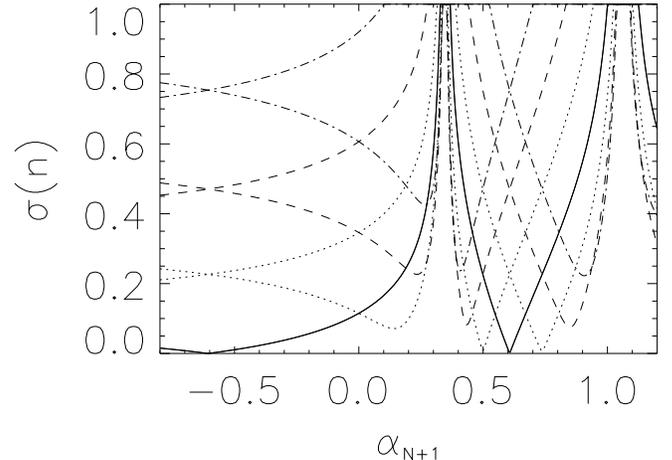}}
\caption[]{\label{fig1}$\sigma(n)$ as a function of $\alpha_{N+1}$
for $N=3$ and $U\Delta \tau=0.05$. Results are shown for 
$n=3$ (solid), $n=2, 4$ (dotted), $n=1, 5$ (dashed) and
$n=0, 6$ (dashed-dotted).  }
\end{minipage}
\hfill
\end{figure}

The free parameter $\alpha_{N+1}$
can now  be used to minimize $\sigma(n)$. For instance, for a half-filled
systems with a large value of $U$, most sites have the occupancy $N$.
It should then be useful to minimize $\sigma(N)$.
Fig. \ref{fig1} shows $\sigma(n)$ as a function of $\alpha_{N+1}$ for 
$N=3$. It illustrates that it is possible to obtain $\sigma(n=3)=0$
and at the same time to also obtain rather small values for $n=2$ and $n=4$.
If the system instead is close to some other integer filling $n$, we can choose
$\alpha_{N+1}$ so that the corresponding $\sigma(n)$ is small.

Since we are often interested in a system at or close to some 
integer occupancy $n_0$,
it is also convenient to introduce
\begin{equation}\label{eq:1a}
e^{-U \Delta \tau \lbrack n(n-1)- n_0(n_0-1)\rbrack /2}=
\sum_{i=k}^{N+1} \tilde w_ke^{-x_k (n-n_0)},
\end{equation}

\noindent

\noindent
\begin{minipage}{3.385in}
\begin{table}
\caption[]{Values of $\tilde w_k$ and $x_k$ 
(Eqs.~(\ref{eq:1a},\ref{eq:1b})) and $\sigma(n)$ (Eq.~(\ref{eq:9})) 
for $N=1$ and $n_0=N$ as a function of $\Delta \tau U$.
 $\tilde w_2=\tilde w_1^{\ast}$ and $x_2=x_1^{\ast}$.
The first lines were obtained by using $K=N+1$ and the following
results by using $K=N$ and minimizing $\sigma(1)(\equiv 0)$.
}       
\begin{tabular}{ccccccc}
 $\Delta \tau U$ & $\tilde w_1$  &  $x_1$   & $\sigma(0)$ & $\sigma(1)$ &
$\sigma(2)$    \\  
\tableline
0.1 & .5+i.0781&  .10-i.3097& 0.501 & 0.156 & 0.156  \\
0.2 & .5+i.1089& .20-i.4290 & 0.701 & 0.212 & 0.212  \\
0.4 & .5+i.1499&.40-i.5823 & 1.194 & 0.300 & 0.300 &  \\
0.1 & .5 & .05-i.31360& 0.324 & 0.000 & 0.324  \\
0.2 & .5 & .1 -i.43980& 0.471 & 0.000 & 0.471  \\
0.4 & .5 & .2 -i.61160& 0.701 & 0.000 & 0.701  \\
\end{tabular}
\label{tableI}
\end{table}
\end{minipage}
where  $\tilde w_k$ is related to $w_k$ and $x_k$
by a simple transformation
\begin{equation}\label{eq:1b}
\tilde w_k=w_k e^{Un_0(n_0-1)/2-x_kn_0}.
\end{equation}

Table \ref{tableI} shows results for $\tilde w_k$, $x_k$ and $\sigma(n)$
for $N=2$ and $n_0=N$. The first lines were obtained by satisfying
Eq.~(\ref{eq:1}) also for the irrelevant case $n=2N+1$, while the
following lines were obtained by satisfying Eq.~(\ref{eq:1})
only for $n\le 2N$ and using the additional freedom to minimize
$\sigma(N)$. In the latter case $\sigma(n)$ is symmetric around
$n=N$ and $\tilde w_k$ (but not $w_k$) is real.

For $N=1$ and $N=2$ we have obtained analytical expressions for 
the coefficients $\tilde w_k$ and $x_k$ which minimize $\sigma(N)$.
Using $n_0=N$ we find the exact transformation for $N=1$
\begin{equation}\label{eq10}
\tilde w_{1,2}=0.5;\hskip0.3cm
x_{1,2}=0.5U\Delta \tau \pm i {\rm tan}^{-1}\sqrt{e^{U\Delta \tau}-1}.
\end{equation}
This result is identical to a result obtained
by Hirsch,\cite{Hirsch} but presented in a different form. For $N=2$ 
we obtain
\begin{eqnarray}\label{eq:11}
&&\tilde w_{1,2}=a;\hskip0.75cm x_{1,2}=
1.5U\Delta \tau \pm i {\rm cos}^{-1}(2\beta-1) \nonumber \\
&&\tilde w_3=1-2a;\hskip0.2cm x_{3}=1.5U\Delta \tau  
\end{eqnarray}
with $\beta=(\gamma^4-1)/(\gamma-1)/4$, $\gamma={\rm exp}(-0.5U\Delta \tau)$
and  $a=(\gamma-1)/(2\beta-2)/2$. We observe that the $\tilde w$'s are 
all real in these cases.

For $N=3$ we have not found a simple analytical expression
for $\tilde w_k$ and $x_k$. Therefore numerical results
are given in Table \ref{tableII} for $n_0=N$. The corresponding values of
$\sigma(n)$ are given in Table \ref{tableIII}.

\noindent
\begin{minipage}{3.385in}
\begin{table}
\caption[]{Values of $\tilde w_k$ and $x_k$ 
for $N=3$ as a function of $\Delta \tau U$ obtained by minimizing
 $\sigma(N)$.                     
($x_2$=$x_1^{\ast}$ and $x_4$=$x_3^{\ast}$).}      
\begin{tabular}{ccccc}
 $\Delta \tau U$ & $\tilde w_1$  &  $\tilde w_3$   & $x_1$  &  $x_3$ \\         
\tableline
0.05& .45154        & .04846   & .1250-i.16385& .125 -i.51551  \\
0.1 & .44890       & .05110   & .25-i.22883 & .25 -i.72002 \\ 
0.2 & .44345       & .05655       & .5-i.31561 & .5-i.99334 \\
0.4 & .43204       &.06796       & 1.-i.42465    & 1.-i1.33776   \\
\end{tabular}
\label{tableII}
\end{table}
\end{minipage}
 
\noindent
\begin{minipage}{3.385in}
\begin{table}
\caption[]{Values of $\sigma(n)$     
for $N=3$ as a function of $\Delta \tau U$ obtained by 
minimizing $\sigma(N)$ and corresponding to the $\tilde w_k$'s 
and $x_k$'s in Table \ref{tableII}.}
\begin{tabular}{cccccccc}
 $\Delta \tau U$ & $\sigma(0)$ & $\sigma(1)$ &  $\sigma(2)$ &
 $\sigma(3)$ & $\sigma(4)$ & $\sigma(5)$ & $\sigma(6)$ \\
\tableline
0.05& 0.75   &0.47    &0.23   & .0000 &0.23   &0.47   &0.75   \\
0.1 & 1.21   &  0.70  & 0.32  & .0000 & 0.32  & 0.70  &1.21   \\ 
0.2 & 2.25   &  1.11  & 0.47  & .0000 & 0.47  & 1.11  &2.25   \\ 
0.4 & 5.97   &  1.99  & 0.70  & .0000 & 0.70  & 1.99  &5.97   \\ 
\end{tabular}
\label{tableIII}
\end{table}
\end{minipage}

So far we have only considered exact transformations of Eq.~(\ref{eq:1}),
which requires a field with $N+1$ values.     Since the sampling procedure
as well as the Trotter decomposition introduce errors, we may also
consider approximate transformations. This has the advantage that we can
then reduce the number of values of the field and make the sampling procedure
more efficient. For this purpose we 
introduce the relative error
\begin{equation}\label{eq:8c}
\Delta(n)={1\over B(n)}\lbrack \sum_{k=1}^{N+1} w_k e^{-x_k n}\rbrack -1.
\end{equation}
We then require that 
\begin{equation}\label{eq:8d}
\sum_{n=0}^{2N}f_n\lbrack |\Delta(n)|+\alpha\sigma(n))\rbrack
\end{equation}
should  be minimized, where $\alpha$ is some appropriate parameter which 
emphasises the suppression of the errors $\Delta$ ($\alpha<<1$)
and $f(n)$ is a weight factor which emphasizes 
the terms close to $n=N$.       
The minimum of Eq.~(\ref{eq:8d}) is searched using a Metropolis algorithm
followed by a Newton minimization.
The precise values of $\tilde w_k$ and $x_k$ depend on the weight
factors $\alpha$ and $f_m$. There is also no guarantee that 
the Metropolis search finds the absolute minimum.
 
In Table~\ref{tableVII} we show results for $N=2$ using a field
with just {\it two} values.     The table shows that if $U\Delta \tau$ 
is not too large, the errors $\Delta$ are small. For instance, for
$U\Delta \tau=0.4$ the largest error is just 0.04, which occurs
for $n=1$ and $n=3$. Such an error means that we are effectively using a
Hamiltonian where $U$ depends on $n$ and has an error of about 4
$\%$ for $n=1$ and $n=3$, but is almost exact for other values
of $n$. It should therefore be possible to treat $N=2$ with
just a single Ising spin, as for $N=1$, although the field in the
$N=2$ is complex in contrast to the real field in one of the Hirsch $N=1$
transformations.\cite{Hirsch} Table~\ref{tableVIII} shows results for $N=3$ and
just {\it two} values of the field. In this case the errors are larger than
in the $N=2$ case, but the larger errors happen for the configurations
with $n=0$ and $n=6$, which should be rare for large values of $U$.

\noindent
\begin{minipage}{3.385in}
\begin{table}
\caption[]{Values of $\tilde w$, $x$, $\Delta$ and $\sigma$
for $N=2$ as a function of $\Delta \tau U$ obtained by 
the expression in Eq.~(\ref{eq:8d}) using just {\it two} values of the field.
$\Delta(n)$ and $\sigma(n)$ are symmetric around $n=2$.}                       

\begin{tabular}{ccccccccc}
 $\Delta \tau U$ & $\tilde w_1$ & $x_1$& $\Delta(0)$ & $\Delta(1)$ & 
 $\Delta(2)$ & $\sigma(0)$ & $\sigma(1)$ & $\sigma(2)$  \\
\tableline
.05& .5 & .075+i.21990 & .0000 & .0006 & .0000& .47 & .22& 0.0 \\
.1 & .5    & .15+i.30580  & .0000  & .0025 & .0000& .70  &.32 &0.0\\ 
.2 & .5 & .3+i.41808   & .0000 & .0100  & .0000& 1.11 & .45&  0.0 \\
.4 & .5 & .6+i.55239   & .0000 & .0397  & .0000& 1.99 & .64&  0.0 \\ 
\end{tabular}
\label{tableVII}
\end{table}
\end{minipage}

\noindent
\begin{minipage}{3.385in}
\begin{table}
\caption[]{Values of $\Delta(i)$     
for $N=3$ as a function of $\Delta \tau U$ obtained by 
the expression in Eq.~(\ref{eq:8d}) using a field with just
{\it two} values of the field.                              
$\Delta(n)$ and $\sigma(n)$ are symmetric around $n=3$.
The values of $\sigma(n)$ are similar as in Table \ref{tableIII}.}                       
\begin{tabular}{ccccccc}
 $\Delta \tau U$ & $\tilde w_1$ & $x_1$ & $\Delta(0)$ & $\Delta(1)$ 
  &  $\Delta(2)$ & $\Delta(3)$  \\
\tableline
0.05& .5     &.125+i.21990 &-.0104 & .0000 & .0006 & .0000 \\
0.1 & .5     &.25+i.30580&-.0466 & .0000 & .0025 & .0000 \\ 
\end{tabular}
\label{tableVIII}
\end{table}
\end{minipage}

In the same spirit we can require that $\tilde w_k$ and $x_k$
are all real. It is still possible to obtain parameters so that
Eq.~(\ref{eq:1}) is rather well satisfied. 
For an equal number of spin up and spin down electrons,
even away from half-filling, 
the spin up and spin determinants are then equal and real and
their product is positive definite. 
Thus the determinants do not cause a sign problem in this case.
For a repulsive $U$ some of the $w_k$ must, however, be negative. 
The sign problem then enters in a different form. It would be 
interesting to test whether or not this sign problem is less serious than in 
other formulations, where it enters via the determinants.

To summarize, we have presented an exact discrete Hubbard-Stratonovich
transformation for a system with the orbital degeneracy $N$ using 
a field which takes $N+1$ complex values. The sign problem in the
traditional Quantum Monte Carlo treatment is therefore converted
into a phase problem. The standard deviation of the transformation 
has been minimized. It will be interesting to see how this influences
the sampling efficiency.  Analytical formulas for the fields were 
provided for $N=1$ and $N=2$ and numerical results for $N=3$. Alternative
approximate transformations were presented, where the number of values  
of the field is smaller than $N+1$. For instance, it was found that for
$N=2$ and $N=3$  and for $\Delta \tau U$ not too large,
a rather accurate transformation can be obtained
with a field which only takes two values, corresponding to an Ising spin.

We would like to thank A. Muramatsu for very stimulating and helpful
discussions.

\end{multicols}
\end{document}